\documentclass[12pt]{article}
\InputIfFileExists{epsf.sty}{}{}
\usepackage{graphicx}
\def\beq{\begin{equation}}
\def\eeq{\end{equation}}
\def\bea{\begin{eqnarray}}
\def\eea{\end{eqnarray}}
\def\nn{\nonumber}
\def\roughly#1{\mathrel{\raise.3ex\hbox
{$#1$\kern-.75em\lower1ex\hbox{$\sim$}}}}

\def\sss{\scriptscriptstyle}

\def\bd{B_d^0}
\def\bdbar{{\overline{B_d^0}}}
\def\bs{B_s^0}
\def\bsbar{{\overline{B_s^0}}}
\def\ks{K_{\sss S}}

\def\btod{{\bar b} \to {\bar d}}
\def\btos{{\bar b} \to {\bar s}}
\def\bra#1{\left\langle #1\right|}
\def\ket#1{\left| #1\right\rangle}
\def\ANPqq{A_{\sss NP}^{qq}}
\def\ANPuu{A_{\sss NP}^{uu}}
\def\ANPdd{A_{\sss NP}^{dd}}
\def\ANPss{A_{\sss NP}^{ss}}

\begin{document}

\begin{flushright}  
UdeM-GPP-TH-04-122 \\
McGill 04/30 \\
\end{flushright}

\begin{center}
\bigskip
{\Large \bf CP Violation in the B System: Measuring New-Physics
Parameters \footnote{talk given at {\it MRST 2004: From Quarks to
Cosmology}, Concordia University, Montreal, May 2004.}} \\
\bigskip
\bigskip
{\large David London\footnote{london@lps.umontreal.ca}}
\end{center}

\begin{center}
{\it Physics Department, McGill University,}\\
{\it 3600 University St., Montr\'eal QC, Canada H3A 2T8}\\
and \\
{\it Laboratoire Ren\'e J.-A. L\'evesque, Universit\'e de Montr\'eal,}\\
{\it C.P. 6128, succ. centre-ville, Montr\'eal, QC,
Canada H3C 3J7} 
\end{center}

\begin{center} 
\bigskip (\today)
\vskip0.5cm
{\Large Abstract\\}
\vskip3truemm
\parbox[t]{\textwidth} {I review CP violation in the standard model
(SM). I also describe the predictions for CP violation in the $B$
system, along with signals for physics beyond the SM. I stress the
numerous contributions of Pat O'Donnell to this subject. Finally, I
discuss a new method for {\it measuring} new-physics parameters in $B$
decays.  This knowledge will allow us to partially identify any new
physics which is found, before its direct production at high-energy
colliders.}
\end{center}

\thispagestyle{empty}
\newpage
\setcounter{page}{1}
\baselineskip=14pt

\section{CP Violation in the Standard Model}

In the standard model (SM), CP violation is due to a complex phase in
the Cabibbo-Kobayashi-Maskawa (CKM) quark mixing matrix \cite{pdg}. A
convenient approximate parametrization, due to Wolfenstein
\cite{Wolfenstein}, follows from the (experimental) fact that one can
write the elements of the CKM matrix in terms of powers of the Cabibbo
angle, $\lambda = 0.22$:
\beq
\left(\begin{array}{ccc}
1-\frac{1}{2}\lambda^2 &  \lambda & A\lambda^3 \left( \rho - i\eta \right) \\
-\lambda & 1-\frac{1}{2}\lambda^2 & A\lambda^2 \\
A\lambda^3\left(1 - \rho - i \eta\right) & -A\lambda^2 & 1
\end{array}\right).
\eeq
It is the appearance of the term $i\eta$ which is responsible for CP
violation. To O($\lambda^3$), this term appears only in $V_{ub}$ and
$V_{td}$.

\begin{figure}
\centerline{\epsfxsize 2.5 truein \epsfbox {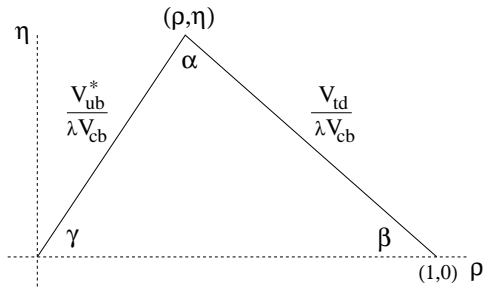}}
\caption{The unitarity triangle.}
\label{Fig1}
\end{figure}

Writing the corner elements as $V_{ub} \equiv |V_{ub}| e^{-i\gamma}$
and $V_{td} \equiv |V_{td}| e^{-i\beta}$, and using the unitarity of
the CKM matrix, $V_{ud} V^*_{ub} + V_{cd} V^*_{cb} + V_{td} V^*_{tb} =
0$, the CKM phase information can be elegantly described by the
so-called unitarity triangle (Fig.~\ref{Fig1}) \cite{pdg}. The
interior angles, $\alpha$, $\beta$ and $\gamma$ describe CP violation
in $B$ decays.  In order to test the SM explanation of CP violation,
the idea is to measure the angles and sides of the unitarity triangle
in as many ways as possible, and to look for consistency. A
discrepancy points to the presence of physics beyond the SM.

\section{CP Violation in the $B$ System}

In general, CP violation requires the interference of two
amplitudes. Consider the decay $B \to f$ and suppose that there are
two amplitudes, with different weak (CP-odd) and strong (CP-even)
phases. The strong phases are typically due to QCD processes, which
are insensitive to whether quarks or antiquarks are involved. I will
return in Sec.~4 to the issue of strong phases. The amplitude for the
CP-conjugate process ${\bar B} \to {\bar f}$ is obtained by simply
reversing the signs of the weak phases. We have
\bea
A(B\to f) & = & A_1 e^{i\phi_1} e^{i\delta_1} + A_2 e^{i\phi_2}
e^{i\delta_2} ~, \nn\\
A({\bar B}\to {\bar f}) & = & A_1 e^{-i\phi_1} e^{i\delta_1} + A_2
e^{-i\phi_2} e^{i\delta_2} ~,
\label{Adiramps}
\eea
where $\phi_{1,2}$ and $\delta_{1,2}$ represent the weak and strong
phases, respectively.

If CP is violated, matter and antimatter behave differently. Thus, CP
violation is signalled by a difference in the rates for the process
and antiprocess. We can therefore define the {\it direct} CP asymmetry
\beq
A_f^{dir} \equiv \frac { \Gamma(B \to f) - \Gamma({\bar B} \to
{\bar f})} {\Gamma(B \to f) + \Gamma({\bar B} \to {\bar f})}
= -\frac{2 A_1 A_2 \sin\Phi \sin\Delta}{A_1^2 + A_2^2 + 2 A_1 A_2
\cos\Phi \cos\Delta} ~,
\label{Adir}
\eeq
where $\Phi \equiv \phi_1 - \phi_2$ and $\Delta \equiv \delta_1 -
\delta_2$. Experimentally, one measures a direct CP asymmetry by
simply comparing the rates for the two processes. Any difference
reflects CP violation. However, recall that the aim is to extract CKM
parameters. From the above expression, we see that the direct CP
asymmetry $A_f^{dir}$ depends on the (unknown) strong phase difference
$\Delta$. Thus, one cannot extract the weak phase information $\Phi$
without hadronic input.

Fortunately, there is another measure of CP violation, which relies on
$B^0$--${\bar B}^0$ mixing. If one chooses a final state $f$ to which
both $B^0$ and ${\bar B}^0$ can decay, then the amplitudes $B^0 \to f$
and $B^0 \to {\bar B}^0 \to f$ will interfere, leading to CP
violation.

In order for this mechanism to produce sizeable effects, large
$B^0$--${\bar B}^0$ mixing is required. Fortunately, in 1987 it was
found that large mixing is present \cite{mixing}. This is arguably the
most important discovery in particle physics in the last 20 years.

The size of this mixing was a great surprise. While it is known that
$\Delta M_d \sim m_t^2$, in 1987 it was expected that $m_t \sim 10$
GeV, which would lead to small mixing. Few people considered the
possibility of large $m_t$. One exception is Ref.~\cite{campodonn}, by
B.~A.~Campbell and P.~J.~O'Donnell. In this paper, various $B$
processes were considered, including mixing, for values $m_t/M_W \le
3.5$. (Experimentally, it is found that $m_t/M_W \sim 2$.)  Thus,
these authors actually {\it anticipated} the large mixing result.

In the presence of large $B^0$--${\bar B}^0$ mixing, one can measure
an {\it indirect} CP asymmetry. There are many final states $f$ which
can be used. The simplest case, which is described below, is where $f$
is a CP eigenstate. Because of mixing, a particle which is ``born'' as
a $B^0$ will quantum-mechanically evolve in time into $B^0(t)$, a
mixture of $B^0$ and ${\bar B}^0$. The measurement of the
time-dependent decay rate then yields two measures of CP violation,
$a_{dir}$ and $a_{indir}$:
\beq
\Gamma(B^0(t) \to f) \sim B + a_{dir} \cos(\Delta M t) + a_{indir} 
\sin(\Delta M t) ~,
\eeq
with
\beq
B \equiv \frac{1}{2} \left( |A|^2 + |{\bar A}|^2 \right) ~,~
a_{dir} \equiv \frac{1}{2} \left( |A|^2 - |{\bar A}|^2 \right)
~,~
a_{indir} \equiv {\rm Im}\left( e^{-2i \phi_{\sss M}} {A}^* {\bar A}
\right) ~,
\eeq
where $\phi_{\sss M}$ is the phase of $B^0$--${\bar B}^0$ mixing. The
quantity $a_{dir}$ is related to the direct CP asymmetry
[Eq.~(\ref{Adir})]. On the other hand, the indirect CP asymmetry
$a_{indir}$ arises due to $B^0$--${\bar B}^0$ mixing. The key point
here, which will be used later, is that the measurement of
$\Gamma(B^0(t) \to f)$ yields 3 observables.

Note that if there is only a single decay amplitude in $B^0 \to f$,
i.e.\ $A_2 = 0$ in Eq.~(\ref{Adiramps}), then $a_{dir} = 0$. However,
we still have $a_{indir} \ne 0$. In fact, this is the most interesting
scenario, since in this case all dependence on the unknown strong
phases vanishes in $a_{indir}$.

Ideally, each of $\alpha$, $\beta$ and $\gamma$ could be measured in
this way. However, although many techniques have been proposed for
getting at the CP phases, only $\beta$ can be measured cleanly through
$a_{indir}$. Here one uses the decay $\bd(t) \to J/\psi \ks$,
dominated (to a very good approximation) by the ${\bar b} \to {\bar c}
c {\bar s}$ tree amplitude, which is proportional to $V_{cb}^* V_{cs}$
and is real. In this case the direct CP asymmetry vanishes, and
indirect CP violation then probes the phase of $\bd$--$\bdbar$ mixing:
$2~{\rm arg}(V_{tb}^* V_{td}) = -2\beta$.

Both BaBar and Belle have measured this CP phase, with the world
average being \cite{HFAG}
\beq
\sin 2\beta = 0.736 \pm 0.049 ~.
\label{sin2beta}
\eeq
As we will see, this agrees with independent measurements.

\begin{figure}
\centerline{\epsfxsize 3.5 truein \epsfbox {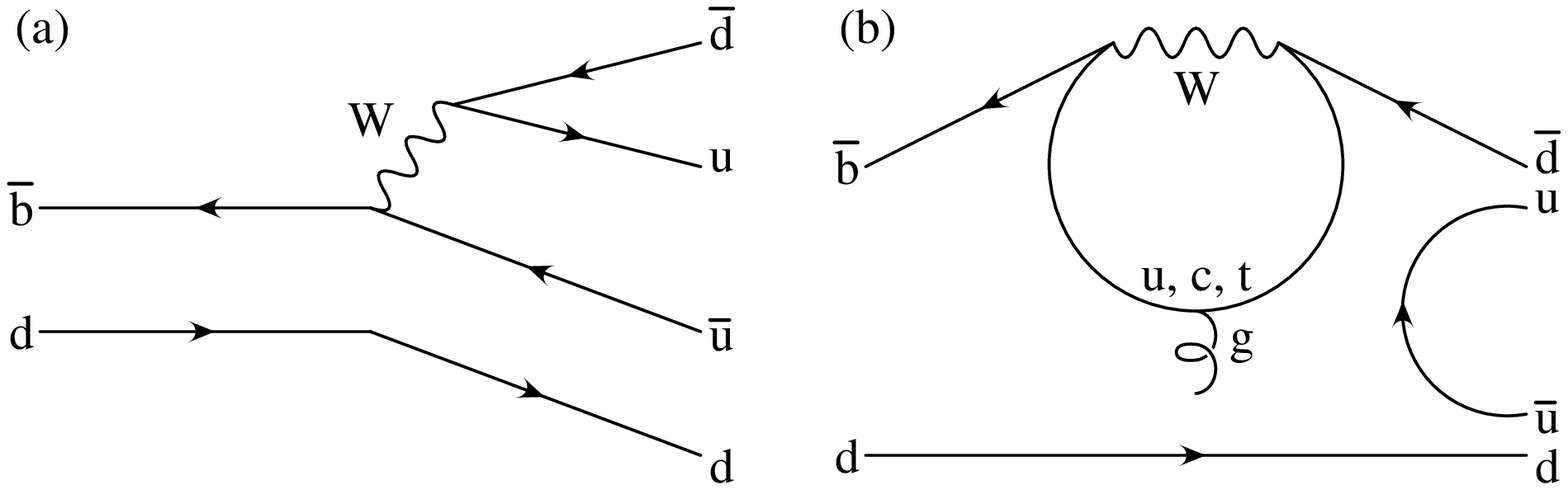}}
\caption{The tree and penguin amplitudes contributing to
$\bd\to\pi^+\pi^-$.}
\label{Fig2}
\end{figure}

The phase $\alpha$ can be extracted from $\bd(t) \to \pi^+\pi^-$. Here
the decay has two contributions (see Fig.~\ref{Fig2}). The tree
diagram ($A_1$) is proportional to $V_{ub}^* V_{ud}$. The penguin
contribution has pieces proportional to $V_{ub}^* V_{ud}$, $V_{cb}^*
V_{cd}$ and $V_{tb}^* V_{td}$. CKM unitarity can be used to write the
$V_{cb}^* V_{cd}$ piece in terms of the other two. The tree amplitude
can then be redefined to include the penguin contribution proportional
to $V_{ub}^* V_{ud}$. The penguin amplitude ($A_2$) can therefore be
taken to be $\sim V_{tb}^* V_{td}$. If the penguin contribution were
zero, the indirect CP asymmetry would probe $2~{\rm arg}(V_{tb}^*
V_{td} V_{ub} V_{ud}^*) = -2(\beta + \gamma) \sim 2 \alpha$.
Unfortunately, $A_2 \ne 0$, i.e.\ the penguins are non-negligible.
Thus, $a_{indir}$ does not probe $\alpha$ cleanly.

Fortunately, a method has been constructed for removing the ``penguin
pollution'' \cite{isospin}. The point is that $\bd\to\pi^+\pi^-$ is
related by isospin to $B^+\to\pi^+\pi^0$ and $\bd\to\pi^0\pi^0$. Using
isospin, the measurement of the branching ratios for
$B^+\to\pi^+\pi^0$ and $\bd\to\pi^0\pi^0$ (and their CP-conjugate
decays) allows us to remove the penguin pollution from
$\Gamma(\bd(t)\to\pi^+\pi^-)$, and obtain $\alpha$ cleanly.

At present, Belle and BaBar have made all of the above measurements
except for the individual $B^0 \to \pi^0 \pi^0$ and ${\bar B}^0 \to
\pi^0\pi^0$ rates. It is not clear exactly when these will be made,
but it is possible that a full isospin analysis will be done, and
$\alpha$ extracted, by the summer of 2005.

For obtaining the angle $\gamma$, many methods have been proposed.
Some of these, such as $B\to DK$ \cite{BDK} have little theoretical
error; others require theoretical input. I will describe one of the
second class of methods. There are two reasons for this. First, I will
come back to this type of technique when discussing the measurement of
new-physics parameters. Second, and more importantly, this method is
being used by BaBar to extract $\gamma$.

\begin{figure}
\centerline{\epsfxsize 3.5 truein \epsfbox {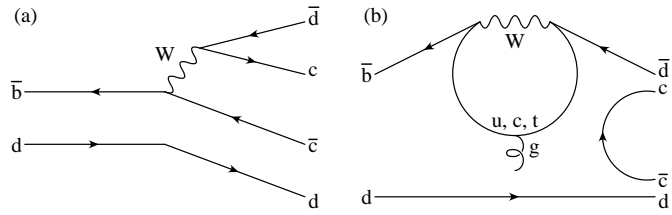}}
\caption{The tree and penguin amplitudes contributing to $\bd\to D^+
D^-$.}
\label{Fig3}
\end{figure}

Consider the decay $\bd(t) \to D^{(*)+} D^{(*)-}$ \cite{BDD}. This is
a ${\bar b} \to {\bar c} c {\bar d}$ transition and has penguin
pollution (see Fig.~\ref{Fig3}). The tree amplitude is $\sim V_{cb}^*
V_{cd}$ and is real. The penguin amplitude can be taken to be $\sim
V_{ub}^* V_{ud}$ after CKM unitarity is applied and the tree redefined
(as in $\bd \to \pi^+\pi^-$). The $\bd \to D^{(*)+} D^{(*)-}$
amplitude can then be written as
\beq
A \sim T e^{i \delta_T} + P e^{i \gamma} e^{i \delta_P} ~,
\label{BDDamp}
\eeq
in which we have explictly written the weak phase $\gamma$ and the
strong phases $\delta_{T,P}$.

It is straightforward to count the number of theoretical
parameters. There are 4: $T$, $P$, $\gamma$ and ${\delta}\equiv
{\delta}_T - {\delta}_P$. (The phase of $\bd$-$\bdbar$ mixing,
$\beta$, is assumed to be measured in $\bd(t) \to J/\psi \ks$.)
However, as noted earlier, the measurement of $\bd(t) \to D^+ D^-$
yields only 3 observables. Thus, in order to extract $\gamma$, it is
necessary to add some theoretical input.

\begin{figure}
\centerline{\epsfxsize 3.5 truein \epsfbox {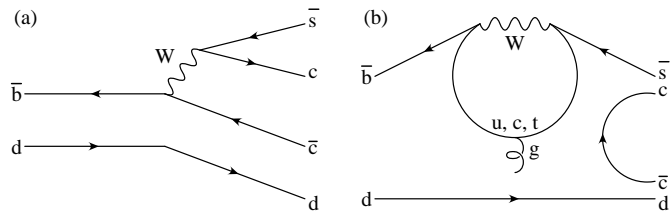}}
\caption{The tree and penguin amplitudes contributing to $\bd\to D_s^+
D^-$.}
\label{Fig4}
\end{figure}

This comes from $\bd \to D_s^{(*)+} D^{(*)-}$, which is a ${\bar b}
\to {\bar c} c {\bar s}$ decay. It also receives both a tree and
penguin contribution (see Fig.~\ref{Fig4}):
\beq
A^{D_s} = T' \, V_{cb}^* V_{cs} + P' \, V_{ub}^* V_{us} \approx
T' e^{i \delta'_T} ~.
\eeq
Here the last approximate equality arises from the fact that
$\left\vert {V_{ub}^* V_{us} / V_{cb}^* V_{cs}} \right\vert \simeq
2\%$. Thus, the decay $\bd \to D_s^{(*)+} D^{(*)-}$ is dominated by
the tree contribution, and the measurement of its rate yields $T'$.

We now make the flavour SU(3) assumption that 
\beq
\frac{\lambda T'}{T} = 1 ~.
\label{su3assump}
\eeq
Given the knowledge of $T'$, this assumption gives us $T$.
Eq.~(\ref{BDDamp}) then contains only 3 theoretical unknowns, and the
3 experimental measurements in $\bd(t) \to D^{(*)+} D^{(*)-}$ can be
used to obtain $\gamma$.

The main theoretical error in this method is the SU(3)-breaking effect
in Eq.~(\ref{su3assump}. The leading-order error is simply given by
the ratio of decay constants, which has been calculated on the lattice
with good precision: $f_{D_s}/f_D = 1.22 \pm 0.04$ \cite{lattice}. The
remaining error is due to second-order effects and is estimated to be
$\sim 10\%$.

As noted above, this method is being used by BaBar to get $\gamma$. It
is possible that we will have a first measurement of $\gamma$ this
summer (2004).

\section{New Physics Signals}

Above, I have described some of the methods used to obtain CKM phase
information. Any discrepancy in the SM description of CP violation
points to the presence of physics beyond the standard model. There are
in fact many signals of such new physics (NP). I list several of these
below.

One test is to compare two $B$ decay modes which in the SM measure the
same CP phase. For example, we know that $\beta$ is obtained in
$\bd(t) \to J/\psi \ks$. But $\beta$ can also be extracted, with
little theoretical error, from pure $\btos$ penguin decays such as
$\bd(t)\to\phi\ks$.

\begin{figure}
\centerline{\epsfxsize 3.5 truein \epsfbox {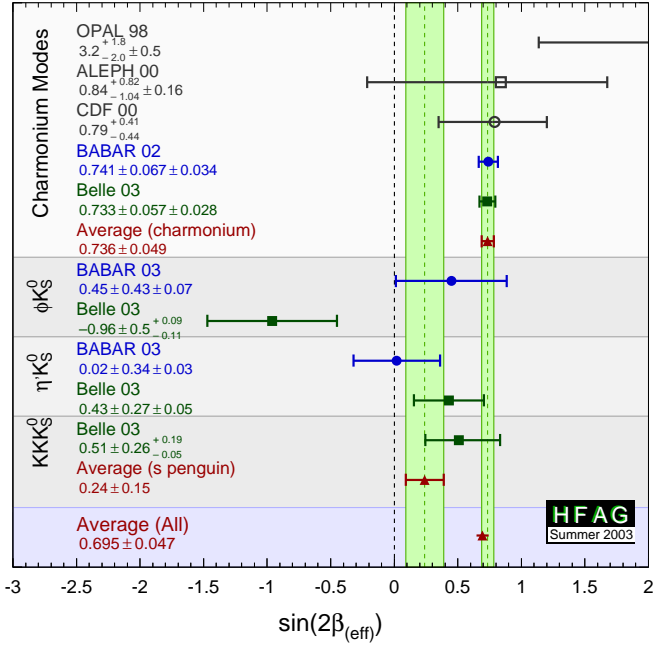}}
\caption{The latest data on the extraction of $\sin 2\beta$ from
measurements of CP violation in $B \to$ charmonium decays and in
$\btos$ penguin decays.}
\label{Fig5}
\end{figure}

The latest data on measurements of $\sin 2\beta$ as extracted from $B$
charmonium decays and $\btos$ penguin decays are shown in
Fig.~\ref{Fig5} \cite{HFAG}. Although the BaBar measurement of $\sin
2\beta$ in $\bd(t) \to\phi\ks$ agrees with that from $\bd(t)\to
J/\psi\ks$ (within errors), Belle finds that $\sin 2\beta = -1$, in
clear disagreement with Eq.~(\ref{sin2beta}). In fact, the value of
$\sin 2\beta$ extracted from all $\btos$ penguin decays is $3.1\sigma$
below that from charmonium decays. Although not yet statistically
significant, this may be pointing to NP.

As noted above, there are in fact many ways to measure CP phases,
often with some theoretical input \cite{Imbeault}. If the values of
the CP angles in these modes disagree with one another at a level
beyond the theoretical error, this implies new physics.

Another hint of new physics comes from $B \to K\pi$ decays. It is
possible to write the amplitudes for these decays in terms of diagrams
($T$, $P$, etc.) \cite{su3}. Some diagrams are expected to be
negligible (e.g.\ exchange- and annihilation-type amplitudes), in
which case we have $R_c = R_n$ \cite{GroRos}, where
\beq
R_c \equiv \frac{2 {\bar\Gamma}(B^+ \to K^+
\pi^0)}{{\bar\Gamma}(B^+\to K^0\pi^+)} ~~,~~~~
R_n \equiv \frac{{\bar\Gamma}(\bd \to K^+ \pi^-)}{2{\bar\Gamma}(\bd\to
K^0\pi^0)} ~.
\eeq
However, present data yields \cite{HFAG}
\beq
R_c = 1.42 \pm 0.18 ~~,~~~~ R_n = 0.89 \pm 0.13 ~.
\eeq
There is a discrepancy of $2.4\sigma$ between $R_c$ and $R_n$, perhaps
suggesting the presence of NP.

There are several observables which are zero (or small) in the SM. For
example, the phase of $\bs$--$\bsbar$ mixing is $2~{\rm arg} (V_{tb}^*
V_{ts}) \simeq 0$. This phase can be measured via the CP asymmetry in
$\bs(t) \to J/\psi \eta$. If this mixing phase is found to be large,
this would indicate NP.

As another example, $BR(B\to \ell^+\ell^-) \sim m_\ell^2/M_B^2$. That
is, this decay is helicity-suppressed, and so its branching ratio is
expected to be tiny. (An exception is $BR(\bs \to \tau^+\tau^-) \sim
10^{-6}$.) These rates were first estimated by B.~A.~Campbell and
P.~J.~O'Donnell in Ref.~\cite{campodonn}. If, for example, $\bs
\to\mu^+\mu^-$ is seen at a measurable level, NP must be present
(e.g.\ SUSY models with large $\tan\beta$).

Since $b\to s\gamma$ is dominated by a single amplitude in the SM, it
is expected that the inclusive $A_{CP}^{dir} (b \to s\gamma) \simeq 0$
\cite{bsgammaref}. This is another good area to search for new
physics. (As an aside, Pat O'Donnell is probably best known for his
calculation of $b\to s\gamma$ in the SM, see
Ref.~\cite{bsgammarate}. These references also discuss other rare
decays, such as $B \to s \ell^+ \ell^-$.)

Another interesting area of study is $B\to V_1 V_2$ decays, where
$V_1$ and $V_2$ are vector mesons. It is possible to measure the
CP-violating triple-product correlation (TP)
${\vec\varepsilon}_1^{*\sss T} \times {\vec\varepsilon}_2^{*\sss T}
\cdot {\hat p}$, where ${\vec\varepsilon}_{1,2}^{*\sss T}$ are the
polarizations of the vector mesons and ${\hat p}$ is the final-state
momentum. In the SM, all TP's are expected to vanish or be very
small\footnote{Note that measurable TP's might be seen in decays
involving radially excited mesons. This was studied by A.~Datta,
H.~J.~Lipkin and P.~J.~O'Donnell in Ref.~\cite{excited}.}, making them
an excellent place to search for new physics \cite{BVVTP}. In fact,
BaBar sees a TP signal in $B \to \phi K^*$ at $1.7\sigma$
\cite{TPsignal}. This is another potential hint of NP.

There are many other examples of this type of signal of new physics.

\begin{figure}
\centerline{\epsfxsize 3.3 truein \epsfbox {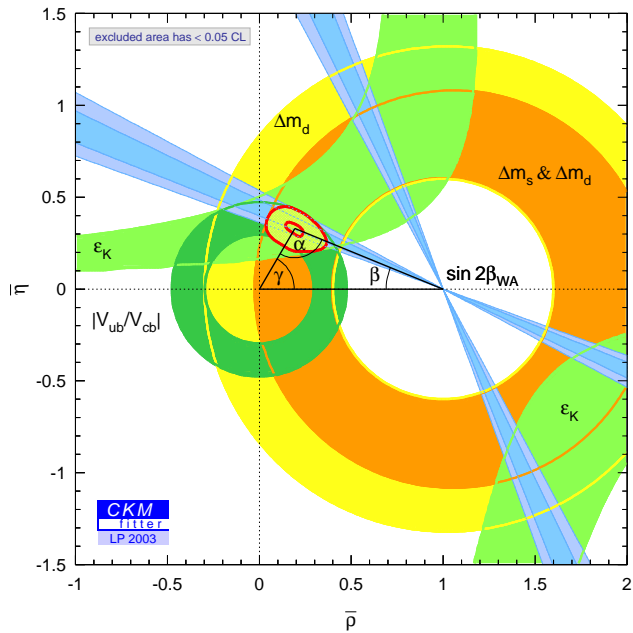}}
\caption{95\% c.l.\ constraints on the unitarity triangle from
independent measurements in the kaon, $\bd$ and $\bs$ systems. The
constraint from $\sin 2\beta$ has not been included in the fit, but is
simply overlaid.}
\label{Fig6}
\end{figure}

Finally, one can search for NP by looking for an inconsistency between
the measurements of the sides and angles of the unitarity
triangle. Fig.~\ref{Fig6} presents the 95\% c.l.\ constraints on the
unitarity triangle coming from independent measurements in the kaon,
$\bd$ and $\bs$ systems \cite{CKMfitter}. As indicated earlier, the
measurement of $\beta$ [Eq.~(\ref{sin2beta})] agrees with that
predicted by these other measurements. The unitarity triangle fit
predicts that the remaining two CP angles should lie in certain
ranges: $78^\circ \le \alpha \le 118^\circ$, $38^\circ \le \gamma \le
79^\circ$ (including hadronic uncertainties). Should either of these
CP phases be found to be outside these ranges, this would imply the
presence of NP.

Note that the measurement of $\sin 2\beta = 0.74$
[Eq.~(\ref{sin2beta})] does not fix $2\beta$: both $2\beta = 48^\circ$
and $132^\circ$ are allowed. In order to test the SM, we need to
distinguish between these two solutions, i.e.\ $\cos 2\beta$ must be
measured \cite{cos2beta}. One possibility, discussed by T.~E.~Browder,
A.~Datta, P.~J.~O'Donnell and S.~Pakvasa \cite{BDDP}, is to study CP
violation in $B \to D^{*+} D^{*-} \ks$ decays. Here both $\sin 2\beta$
and $\cos 2\beta$ can be obtained with some theoretical input
\cite{Datta}.

The bottom line is that there are {\it many} ways of looking for
physics beyond the SM through CP violation in the $B$ system. (Another
method is discussed in Ref.~\cite{VPage}.)

\section{Measuring New-Physics Parameters}

Suppose that some signal of physics beyond the SM is found in $B$
decays. Having confirmed the presence of new physics, we will want to
identify it. Until recently, it was thought that this would have to
wait for a high-energy collider such as the LHC, where the new
particles can be produced directly. However, this is not necessarily
true. As I will show below, it is possible to {\it measure} the NP
parameters through CP violation in $B$ decays \cite{NPmeas}. This
knowledge will allow us to exclude certain NP models and will permit a
partial identification of the NP, before the LHC.

New physics enters principally in loops in $B$ processes. In general,
if NP is present in $\bd$--$\bdbar$ mixing, it will also affect
$\btod$ penguin amplitudes (and similarly for $\bs$--$\bsbar$ mixing
and $\btos$ penguins). As discussed in the previous section, at
present there are several hints of new physics.  All of these hints
occur in processes involving $\btos$ penguin diagrams --- the $\btod$
penguins appear to be unaffected. In line with this, we assume that NP
is present only in $\btos$ transitions.  Furthermore, in order that
the effects be measurable, we assume that the NP operators are roughly
the same size as the SM penguin operators.

Assuming that the new physics affects $\btos$ penguin transitions, it
will lead to new effective $\btos q {\bar q}$ operators. There are 20
possible NP operators $O_{\sss NP}^{ij,q} \sim \bar{s} \Gamma_i b \,
\bar{q} \Gamma_j q$ (here $\Gamma_{i,j}$ represent Lorentz structures;
colour indices are suppressed). In general, there can be new weak and
strong phases associated with each operator. A priori, the question of
which operators are present is model-dependent. Performing a
model-independent analysis therefore appears to be an intractable
mess. Fortunately, it is possible to simplify things.

To see this, it is important to look at strong phases in more
detail. In Sec.~2, I noted that strong phases are due to QCD
processes. In particular, these phases are generated by rescattering.
For example, in the SM, the strong phases of $\btos s {\bar s}$
operators come principally from rescattering from the tree-level
$\btos c {\bar c}$ operators which have large CKM matrix elements.
However, whereas the tree operator has Wilson coefficient $\sim 1$,
the largest rescattered penguin operator has Wilson coefficient $\sim
0.05$. That is, the rescattered amplitude is $\sim 5\%$ as large as
the amplitude causing the rescattering.

The new-physics strong phases are generated by rescattering from NP
operators. However, the NP operators are only expected to be about as
big as SM penguins. Thus, rescattered NP operators are only $\sim 5\%$
as large as this, which is quite small. It is therefore a reasonable
approximation to neglect all NP rescattering. This implies that the NP
strong phases are negligible relative to the SM strong phases.

The neglect of new-physics strong phases leads to a great
simplification. The NP contributes to the decay $B\to f$ through the
matrix elements $\bra{f} O_{\sss NP}^{ij,q} \ket{B}$. We denote each
of the 20 NP $\btos q {\bar q}$ matrix elements by $A_i
e^{i\phi_i^{qq}}$, where $\phi_i^{qq}$ is the weak phase. The point is
that we can now {\it sum} all of the contributions into a single
effective NP matrix element: 
\beq
\sum A_i e^{i\phi_i^{qq}} = \ANPqq e^{i \Phi_{qq}} ~.
\eeq
In other words, for a given $b\to s {\bar q} q$ process
($q=u,d,s,c$), the effects of all NP operators can be parametrized in
terms of a single effective NP amplitude $\ANPqq$ and the
corresponding weak phase $\Phi_{qq}$.

Furthermore, these NP parameters can be {\it measured}. Their
knowledge will allow us to rule out many NP models, giving us a
partial identification before LHC. There are several ways to make such
measurements. I present one such method below. (It is similar to that
used to extract $\gamma$ in the SM from $\bd(t) \to D^{(*)+} D^{(*)-}$
and $\bd \to D_s^{(*)+} D^{(*)-}$, described in Sec.~2.)

Consider the decay $\bd(t)\to\phi\ks$. In the SM, there is a single
decay amplitude. This is a $\btos s{\bar s}$ penguin, whose weak
phase is zero. In the presence of NP, its amplitude can be written
\beq
A' = P'_{\sss SM} e^{i \delta'_{\sss P}} + \ANPss e^{i\Phi_{ss}} ~.
\eeq
Assuming that the phase of $\bd$--$\bdbar$ mixing ($\beta$) is known
independently, the above amplitude is described by 4 theoretical
parameters: $P'_{\sss SM}$, $\ANPss$, $\delta'_{\sss P}$ and
$\Phi_{ss}$. However, the measurement of the time-dependent rate for
$\bd(t)\to\phi\ks$ yields only 3 observables. We therefore need to add
theoretical input in order to extract all theoretical parameters.

This input comes from $\bs(t) \to \phi\ks$. This is a $\btod s {\bar
s}$ penguin decay, and therefore has no NP contributions. Its
amplitude is
\beq
A = P_u e^{i\gamma} e^{i \delta_u} + P_{\sss SM} e^{i \delta_P} ~.
\eeq
Here too there are 3 observables and 4 theoretical parameters.
However, if we assume that $\gamma$ is known from independent
non-$\btos$ measurements (e.g.\ $B \to DK$ decays), the measurement of
this time-dependent rate allows us to extract $P_{\sss SM}$, $P_u$ and
$\delta\equiv\delta_u-\delta_P$. We can then obtain $P'_{\sss SM}$
using the SU(3) relation $\lambda P'_{\sss SM}/P_{\sss SM} = 1$. With
this input, we can then extract $\ANPss$ and $\Phi_{ss}$.

To summarize: the above method allows us to measure the NP parameters
$\ANPss$ and $\Phi_{ss}$. In fact, it is possible to obtain all
$\ANPqq$ and $\Phi_{qq}$ ($q=u,d,s,c$) similarly. (There are other
methods as well which can be used to make such measurements
\cite{DILPSS}.) This knowledge will allow us to distinguish among
possible NP models. For example, some models (e.g.\ gluonic penguin
operators with an enhanced chromomagnetic moment \cite{chromo})
conserve isospin. In such models, one has $\ANPuu = \ANPdd$. Other
models (e.g.\ $Z$- and $Z'$-mediated flavour-changing neutral currents
\cite{ZFCNC}) predict that the NP phase $\Phi_{qq}$ is
universal. Finally, in general, the values of $\ANPqq$ and $\Phi_{qq}$
found are process-dependent.  However, some models (e.g.\ SUSY with
R-parity breaking) predict that NP contributions to certain $\btos
q{\bar q}$ decays are process-independent. By measuring the NP
parameters, all of these predictions can be tested.  In this way we
can exclude certain classes of NP models.

The bottom line is that, assuming that new physics is discovered
through measurements of CP violation in the $B$ system, one can {\it
measure} the NP parameters $\ANPqq$ and $\Phi_{qq}$. Their knowledge
will allow a partial identification of the NP, before its direct
production at LHC.

\section{Conclusions}

To summarize: the raison d'\^etre of $B$ physics is to find physics
beyond the SM. There are {\it many} signals of new physics in
measurements of CP violation in $B$ decays. Given a NP signal, it is
possible to {\it measure} the NP parameters. This will allow a partial
identification of the NP, before direct measurements at future
high-energy colliders. Hopefully, we will find evidence of NP at $B$
factories, measure its parameters and (partially) identify it. Pat
O'Donnell has contributed significantly to this endeavour.

\section*{Acknowledgments}

I thank A. Datta for collaboration on several of the topics discussed
in this talk. This work was financially supported by NSERC of Canada.

\end{document}